\documentstyle[12pt]{article}
\begin{document} 

\title{The Time Evolution of Quantum
Universe in the Quantum Potential Picture} 
\author{A.\ B\l{}aut\thanks{e-mail address 
ablaut@ift.uni.wroc.pl} and J.\ Kowalski--Glikman\thanks{e-mail 
address 
jurekk@ift.uni.wroc.pl}\\ 
Institute for Theoretical Physics\\ 
University of Wroc\l{}aw\\ 
Pl.\ Maxa Borna 9\\ 
Pl--50-204 Wroc\l{}aw, Poland} 
\maketitle 
 
\begin{abstract} 
We use the quantum potential approach to analyse the quantum 
cosmological model of the universe. The quantum potential arises from 
      exact solutions of the full Wheeler-De Witt equation.
 
\end{abstract} 
\vspace{12pt} 
PACT number 04.60 Ds 
\clearpage

\section{Introduction} 
In the recent paper \cite{aj}
a class of solution of the regularized
Wheeler-De Witt equation was given.
An interpretation of the resulting "wave functionals of the
universe" in terms of the modified field dynamics
was also proposed, namely, 
in the properly extended version of the quantum potential 
approach to the quantum mechanics given by David Bohm (see e.g.
\cite{bohm},\cite{jbohm},\cite{shtanov}). In this approach the
time evolution of the fields is generated by the
Hamiltonian of the classical system modified by the additional term,
the so called quantum potential.
The purpose of the present note is to use the above language
to analyse a simple quantum cosmological model.

In the second section we shortly repeat the most important steps 
leading 
to the formulation of the quantum theory, essentialy quantum gravity, 
in 
the quantum potential language. We refer the reader interested in the 
details to the paper \cite{jbohm}.
In what follows we use the notation of \cite{aj}.
As compared with the work \cite{jbohm} here we have to do  with one
important modification resulting from the presence of the $L_{ab}$ 
term 
in our Wheeler - De Witt operator.

In the third section we will find the equation of motion governing
the simplified cosmological model, the homogeneous and isotropic
universe. It turns that it has no singular points and the scale 
factor
grows exponentialy near the classical singularity $a=0$. 

\section{Quantum potential interpretation} 

We recall that the regularized Wheeler-De Witt
operator used in \cite{aj} had the form
$$
{\cal H} =
-\kappa^2\hbar^2\int dx' K(x,x';t)G_{abcd}(x')\frac{\delta}{\delta
h_{ab}(x)}\frac{\delta}{\delta h_{cd}(x')}+
$$
\begin{equation}
+\kappa^2\hbar^2L_{ab}(x)\frac{\delta}{\delta h_{ab}(x)}+
\frac1{\kappa^2}
\sqrt{h(x)} 
(R(x)+2\Lambda),
\label{qham}
\end{equation}
where the function $K(x,x';t)$ responsible for the point-splitting
and the function $L_{ab}(x)$ corresponding to the operator ordering
were fixed during the process of regularization and renormalization.
The renormalized action of ${\cal H}$ on the states
was also defined.
Our task is to apply the quantum potential approach to the 
Wheeler-De Witt equation with quantum Hamiltonian (\ref{qham}).
We assume that the wave function of the universe is of the form
\begin{equation} 
\Psi=e^{\Gamma}e^{i\Sigma}, 
\end{equation} 
with $\Gamma$ and $\Sigma$ the real functionals of the three-metric.
Substituting this to the Wheeler--De Witt
equation and taking the real part\footnote{In this
paper we do not consider the imaginary 
part.
For its interpretation see for example \cite{bohm}.}
we obtain
$$ 
-\kappa^2G_{abcd}(x)\frac{\delta\Sigma}{\delta 
h_{ab}(x)}\frac{\delta\Sigma}{\delta h_{cd}(x)}+\frac1{\kappa^2} 
\sqrt{h(x)} 
(R(x)+2\Lambda) 
+
$$ 
\begin{equation} 
+L_{ab}(x)\frac{\delta\Gamma}{\delta h_{ab}(x)} 
+e^{-\Gamma} 
\kappa^2\left(\frac{\delta^{2}e^{\Gamma}}{\delta h^{2}}%
\right)_{ren}(x)=0,
\label{qhj}
\end{equation}
where in the last term we used the abbreviated notation to 
indicate that 
the action of the second functional derivative is renormalized.
Then we define the
momenta as the functional gradient of $\Sigma$,
to wit 
\begin{equation} 
p^{ab}(x) = \frac{\delta\Sigma}{\delta h_{ab}(x)}.\label{p} 
\end{equation}
With this identification (\ref{qhj}) turns to the Hamilton-Jacobi
equation for general 
relativity
with additional term corresponding to quantum potential:
$$
\kappa^2G_{abcd}(x)p^{ab}(x)p^{cd}(x)
-\frac1{\kappa^2}
\sqrt{h(x)} 
(R(x)+2\Lambda) 
+
$$ 
\begin{equation} 
-\hbar^2L_{ab}(x)\frac{\delta\Gamma}{\delta h_{ab}(x)}
-\hbar^2e^{-\Gamma}
\kappa^2\left(\frac{\delta^{2}e^{\Gamma}}{\delta h^{2}}%
\right)_{ren}(x)=0.
\label{HJ}
\end{equation}
We see that in the limit $\hbar\rightarrow 0$ we obtain the
classical hamiltonian constraint.
The wave function is subject to the second set of equations, 
namely the
ones enforcing the three dimensional diffeomorphism invariance. 
These 
equations read (for imaginary part) 
\begin{equation} 
\nabla^a\frac{\delta\Sigma}{\delta h_{ab}(x)} =
 \nabla^a\, p_{ab} =0 
\label{3diff} 
\end{equation} 
Thus our theory is defined by two equations (\ref{HJ})
and
(\ref{3diff}). Now we can follow without any alternations the 
derivation 
of Gerlach \cite{gerlach} to obtain the full set of ten equations 
governing the quantum gravity theory in the quantum potential 
approach 
\begin{eqnarray} 
0&=& {\cal H}^a = \nabla_a\, p^{ab} ,\label{diff1}\\ 
0&=& {\cal H}_{\bot} = 
-\kappa^2G_{abcd}(x)p^{ab}p^{cd}+\frac1{\kappa^2}\sqrt{h(x)} 
(R(x)+2\Lambda)\nonumber \\ 
&+&L_{ab}(x)\frac{\delta\Gamma}{\delta h_{ab}(x)} 
+\kappa^2e^{-\Gamma} 
\left(\frac{\delta^{2}\,e^{\Gamma}}{\delta h^{2}}\right)_{ren}(x), 
\label{35}\\ 
&&\dot{h}_{ab}(x,t) = \left\{ h_{ab}(x,t),\, {\cal H}[N,\vec{N}]%
\right\},\label{1}\\ 
&&\dot{p}^{ab}(x,t) = \left\{ p^{ab}(x,t),\, {\cal H}[N,\vec{N}]%
\right\}.\label{2} 
\end{eqnarray} 
In equations above, $\{\star,\, \star\}$ is the usual Poisson 
bracket, and 
\begin{equation} 
{\cal H}[N,\vec{N}]=\int\, d^3x \left( N(x){\cal H}_{\bot}(x) + 
N^a(x){\cal 
H}_a(x)\right) 
\label{fullh}
\end{equation} 
is the total hamiltonian (which is a combination of constraints).
The above set of equations describes the time evolution corresponding
to  a given solution of the Wheeler-De Witt equation
of the classical fields $h_{ab}(t)$ and $p^{ab}(t)$ from given 
initial
surface. In this way, in this approach we are able to circumvent the
problem of time of quantum gravity \cite{isham}.
It should be noted that because of the presense of the quantum 
potential
term in the superpotential the algebra
of
${\cal H}$  doesn't close in general case, 
so it must be checked
for every solution separately.

It 
might seem puzzling at the first sight why to a single wavefunction 
there corresponds a set of equations with, clearly, many solutions. 
The 
resolution of this problem is that the wavefunction, as a rule, is 
sensitive only to some aspects of the configuration. For example, the 
exact wavefunction $\Psi_I=exp(-\frac{3\rho^{(5)}}{\Lambda}{\cal V})$
found in \cite{aj}
depends on the total volume of the universe, ${\cal V}$, only,
and thus any 
configuration with given volume leads to the same numerical value 
of it. 
The above dynamical equations provide us with much more detailed 
information concerning the dynamics of the system than the 
wavefunction 
alone. 

\section{ A simple model}

In \cite{aj} we have found three real solutions of the
 Wheeler-De Witt equation.
Such states were interpreted as "frozen in time"
since real solutions do not, by definition, evolve in time 
(cf.\cite{jbohm}).
We therefore take the complex superposition of the solutions,
\begin{equation}
\Psi= a \Psi_I + b \Psi_{II},
\label{cs}
\end{equation}
$$
a=|a|e^{i\alpha}, b=|b|e^{i\beta}\;\;\;\;|a|=|b|,
$$
where $\Psi_I=exp(-\frac{3\rho^{(5)}}{\Lambda}{\cal V})$, $\Psi_{II}
=exp(\frac43\Lambda\frac{1}{\kappa^4\hbar^2\rho^{(5)}}{\cal R})$ are 
two
exact solutions of the Wheeler--De Witt equation found in \cite{aj}.
Here
${\cal V}=\int \sqrt{h}$ is the volume of the universe,
${\cal R}=\int\sqrt{h}R^{(3)}$ its average curvature, and 
$\rho^{(5)}$
is the renormalization constant.
Now we can follow the prescription given in the previous section
to get the dynamical equation for our system corresponding to the
state (\ref{cs}).

The hamiltonian (\ref{35}) can be computed to be
$$
H_\bot = \kappa^2 G_{abcd}\pi^{ab}\pi^{cd} +$$
\begin{equation} 
\label{qhA}
\frac{A \Psi_I^2\Psi_{II}^2}{|\Psi|^4}
\left(
\frac{27}{16}\frac{\rho^{(5)}{}^2\hbar^2\kappa^2}{\Lambda^2} \sqrt h
+\frac{1}{\kappa^2} \sqrt h R -
\frac{8}{9} \frac{\Lambda^2}{\hbar^2\kappa^6\rho^{(5)}{}^2}
\sqrt h \left( -\frac38 R^2 + R_{ab}R^{ab}\right)\right),
\end{equation}
where $A = 2 |a|^4 \sin^2(\alpha-\beta)$ is a parameter which measures
the rate of mixture of two universes.

We can readily write down the dynamical equations of motion (in the
gauge $\vec{N}=0$ and $N=N(t)$.)
Equation (\ref{1}) takes the form
\begin{equation}
\dot h_{ab} = 2 N \kappa^2 G_{abcd}\pi^{cd},
\end{equation}
which can be solved for $\pi^{ab}$ in  a standard way
\begin{equation}
\pi^{ab}= \frac{\sqrt h}{2 \kappa^2 N}\left(\dot h^{ab}-
\mbox{tr} (\dot h)h^{ab}\right),
\end{equation}
where $\dot h^{ab} = h^{ac}h^{bd}\dot h_{cd}$,
$\mbox{tr} (\dot h)= h^{ab}\dot h_{ab}$.
Equation (\ref{2}) takes the form
$$
\dot\pi^{ab} = \frac{1}{2\kappa^2 N}\left( \ddot h_{cd} G^{abcd}+
\dot h_{cd} \dot G^{abcd}- \dot h_{cd} G^{abcd}\frac{\dot 
N}{N}\right) =
$$
$$
= \frac{\sqrt h}{N\kappa^2}\left[
\frac18 h^{ab} \mbox{tr} (\dot h \times \dot h) - \frac12
(\dot h \times \dot h)^{ab} + \frac12 \dot h^{ab}\mbox{tr} (\dot h)
\right]
$$
$$
- \frac{N\sqrt h}{\kappa^2}{\cal F}\left( \frac12 h^{ab}R - 
R^{ab}\right)
-\frac{27N\sqrt h}{32}\frac{\rho^{(5)}{}^2\hbar^2\kappa^2}{\Lambda^2} 
{\cal
F}h^{ab}
$$
$$
+\frac{8 N \sqrt h}{9} \frac{\Lambda^2}{\hbar^2\kappa^6
\rho^{(5)}{}^2}
{\cal F}\left[\frac12 h^{ab} R_{cd}R^{cd} -4 R^{c(a}R_c^{b)} -
\frac{3}{16}
R^2 h^{ab} + \frac34 R R^{ab} +\right.
$$
$$
+ \left. \frac12 \nabla^{ab} R - \Box R^{ab} + \frac14 h^{ab} \Box R
\right]
$$
\begin{equation}
- N \int
\sqrt h \left[
\frac{27}{16}\frac{\rho^{(5)}{}^2\hbar^2\kappa^2}{\Lambda^2}
+ \frac{1}{\kappa^2} R
- \frac{8}{9} \frac{\Lambda^2}{\hbar^2\kappa^6\rho^{(5)}{}^2}
\left( -\frac38 R^2 + R_{ab}R^{ab}\right)
\right] \frac{\delta {\cal F}}{\delta h_{ab}},\label{eqmQP}
\end{equation}
where
\begin{equation}
{\cal F} =
\frac12%
\frac{\sin^2(\alpha -\beta)}{%
\left\{ \cosh\left(  \frac{3 \rho^{(5)}}{\Lambda} {\cal V} +
\frac{4\Lambda}{3 \hbar^2\kappa^4 \rho^{(5)}}{\cal R}\right)
 + \cos(\alpha -\beta)
\right\}^{2}}
\end{equation}

We see that quantum effects exhibit themselves in two ways: first 
there
are higher curvature terms in the effective quantum hamiltonian, and
second the resulting coupling constants are modified not only by 
quantum
corrections following from the renormalization, but also by nonlocal
terms related to the globasl structure of the universe. This latter 
fact
was to be expected in the framework based on the quantum potential
approach since the quantum potential is usually nonlocal, however it
cannot be merely treated as a mere artefact of the method employed. 
Even
if the ``reality'' of the field evolution described by equations
(\ref{35}), (\ref{1}), (\ref{2}) can be questioned in the highly 
quantum
regime, without doubts these equations provide a correct semiclassical
approximations, and in this approximation some traces of nonlocality
will still be present. It is our opinion however that the nonlocal 
terms
discovered above do reflect the deep structure of quantum gravity. 
These
questions and the analysis of solutions of the dynamical equation in 
the
cosmological context will be subject of the separate paper.
\newline

Now we present a solution of the resulting equations
assuming that
$R_{ab}=0$,
$h_{ab}(t)=a(t){\tilde h}_{ab}$, where $\tilde h$ is a flat metric on
compact three-manifold (space) 
with normalization $\int \sqrt{\tilde h}=1$.
The equation of motion for this ansatz that follow from
(\ref{eqmQP}) reduce to the 
following equation:
$$
\frac{\ddot a}{a N^2}-\frac14\frac{{\dot a}^2}{a^2 N^2}
-\frac{\dot a}{a}\frac{{\dot N}}{N^3}+
$$
\begin{equation}
\label{aem}
-B\frac{\sin^2(\alpha-\beta)}{\left[ 
\cosh(\frac{3\rho^{(5)}}{\Lambda}a^{3/2})+\cos(\alpha-\beta) 
\right]^3}\left[ 
\frac14-\frac{3\rho^{(5)}}{2\Lambda}a^{3/2}
\sinh(\frac{3\rho^{(5)}}{\Lambda}a^{3/2}) \right]=0,
\end{equation}
where
$\frac{B}{\kappa^2}=\frac{27}{16}
\frac{{\rho^{(5)}}^2 \hbar^2 \kappa^2}{\Lambda^2}$.
In agreement with what was mentioned above, we see that the limit 
$\hbar\rightarrow 0$ 
corresponds to the classical model. 
The Hamiltonian of the system which is the reduced version of 
(\ref{qhA}) has the form
\begin{equation}
\label{aham}
H(x)=\frac{1}{\kappa^2}\left( - \frac32 \frac{{\dot a}^2}{a^{1/2}N^2}
+\frac{B}{2} 
a^{3/2}\frac{\sin^2(\alpha-\beta)}
{\left[\cosh(\frac{3\rho^{(5)}}{\Lambda}a^{3/2})
+\cos(\alpha-\beta)\right]}\right).
\end{equation}
It can be easily checked that the evolution (\ref{aem})
is generated by the Hamiltonian (\ref{aham}).
It also follows from the form of (\ref{aham}) that after  
reduction we are left with the reparametrization invariant theory.
The solution of (\ref{aham}) near the point
$a=0$ are modified as compared to their classical singular
behaviour because of the quantum potential term that acts effectively 
as a nonzero cosmological constant in that region.
The scale factor grows here exponentialy
\begin{equation}
a\sim 
\exp\left(\pm\frac34
\frac{\kappa^2\hbar\rho^{(5)}}{\Lambda}t\right).
\end{equation}
For $a \gg 1$ quantum potential vanishes and we come back
to the classical case.

\section{Conclusions} 
 
In the paper we analysed time evolution of quantum universes that 
arises
as a complex combination of exact solutions of the Wheeler--De Witt
equation presented in \cite{aj}
 by making
use of the quantum 
potential interperetation. 
We noticed that
in the simplified situation considered explicitly,
 the theory is reparametrization
invariant contrary to the general case were the quantum potential may
spoil the invariance in  time direction. We show that in the 
simplified
model considered above the initial cosmic singularity is avoided. It 
is
a matter of future investigations to check as to whether this property
is generic for quantum universes described by equations (\ref{eqmQP})
 or not.

\setcounter{equation}{0}
\renewcommand{\theequation}{A.\arabic{equation}}

\end{document}